\documentstyle[12pt,aas2pp4]{article}

\received{..}
\accepted{..}
\journalid{The Astrophysical Journal Letters}{}

\lefthead{Moll\'{a} \& Roy}
\righthead{Modeling the Galaxy NGC~1313 }

\def\msun{\thinspace\hbox{$\hbox{M}_{\odot}$}}

\begin{document}

\setcounter{page}{1}
\def\counter{1}

\title{MODELING THE RADIAL ABUNDANCE DISTRIBUTION OF THE TRANSITION
GALAXY NGC~1313}

\author{Mercedes Moll\'a and Jean-Ren\'e Roy}

\affil{Département de Physique and Observatoire du mont Mégantic,
 Universit\'{e} Laval, Québec, G1K 7P4 Qc, Canada}

%\offprints{M. Moll\'{a}}

%\date{Received xxxx 1998; accepted xxxx 1998}
\today


%\maketitle

\begin{abstract}

NGC~1313 is the most massive disk galaxy showing a flat radial
abundance distribution in its interstellar gas, a behavior generally
observed in magellanic and irregular galaxies. We have attempted to
reproduce this flat abundance distribution using a multiphase chemical
evolution model, which has been previously used successfully to depict
other spiral galaxies along the Hubble morphological sequence.  We
found that it is not possible to reproduce the flat radial abundance
distribution in NGC~1313, and at the same time, be consistent with
observed radial distributions of other key parameters such as the
surface gas density and star formation profiles.  We conclude that a
more complicated galactic evolution model including radial flows, and
possibly mass loss due to supernova explosions and winds, is necessary
to explain the apparent chemical uniformity of the disk of NGC~1313.

\end{abstract}

\section{INTRODUCTION}

NGC~1313 is often described as a galaxy in transition between the
magellanic type galaxies and normal disk galaxies such as the Sc
galaxies M~33 (NGC~598) and NGC~300. Low-mass galaxies have maximum
rotation velocity between 40 and 70 km/s, (Carignan \& Freeman 1985;
Carignan \& Beaulieu 1989; Coté, Freeman, \& Carignan 1997), while
normal galaxies of latest morphological types have values ranging from
90 to 110 km/s (Puche, Carignan, \& Bosma 1990; Carignan \& Puche
1990; Zaritsky et al. 1994). Chemically, Sc and Sd galaxies show very
steep radial distributions of abundances (see e.g. Zaritsky et
al. 1994), while low-mass irregulars have very homogeneous abundance
distributions (Pagel et al. 1978; Kobulnicky \& Skillman 1996; Roy et
al. 1996; Devost, Roy, \& Drissen 1997; Kobulnicky 1998).  The galaxy
NGC~1313 has a maximum rotation velocity very similar to that of
NGC~300 and NGC~598. However, contrary to them, its radial
distribution of oxygen abundance appears very uniform, with a mean
value around 12 $+$ log (O/H) = 8.4, a level similar to what is
observed in the Large Magellanic Cloud (Pagel et al. 1980; Walsh \&
Roy 1997).  Thus NGC~1313 is the most massive disk galaxy known
showing very little or no radial gradient in its radial abundance
distribution.

Low-mass galaxies are usually irregular in their optical appearance
due to the smaller mass, and consequently, the smaller potential
prevents the formation of spiral structure. These galaxies have large
relative quantities of atomic gas; because molecular gas is difficult
to detect in these systems, it is not yet understood if this is truely
due to low content. The atomic gas distribution shows an exponential
decrease with galactocentric radius from a central maximum. Moreover,
many galaxies are undergoing a burst of star formation in the center
(Greggio et al. 1992; Gallart et al. 1996; Skillman, Dohm-Palmer, \&
Kobulnicky 1998); they have low metallicities (Garnett et al. 1995).
These facts and the intrinsic blue colors (Skillman 1996; Pildis,
Schombert, \& Eder 1997) indicate that low-mass galaxies are
relatively unevolved.

The rotation curve can generally be measured far from the galactic
center, and the velocity behavior suggests a large proportion of dark
matter in these galaxies. However an alternate explanation within the
framework of the MOND theory of gravitation, which gives flat rotation
curves for the observable matter in galaxies, is possible (McGaugh \&
de Block 1998).

The systematic variation of abundances with galactocentric radius in
normal spiral galaxies has been known for a long time (see references
in D\'{\i}az 1989; Skillman 1997; Garnett et al. 1997; Henry
1998). Radial abundance gradients vary in amplitude.  Correlations
between their amplitudes and other galaxy characteristics have been
clearly established (Vila-Costas \& Edmunds 1992; Zaritsky et
al. 1994; Martin \& Roy 1994). The radial gradient of abundances is
steeper in later types of galaxies and the mean abundance level is
moderate. For early type galaxies, gradients appear flatter and mean
abundances seem higher (Oey \& Kennicutt 1993), but this is based on
small samples due to the difficulty of measuring the O/H abundances in
these systems.  Gas distributions in normal galaxies also have some
trends: the atomic gas shows an apparent hole in the central region,
while the maximum column density is reached further away into the
disk. The position of this maximum is shifted towards the inner disk
for late type galaxies, while for the early types, the maximum is
further out in the disk (Moll\'{a}, Ferrini, \& D\'\i az 1997; Broeils
\& Rhee 1997).  The proportion of gas in molecular or atomic
phases also depends on the Hubble type, with the larger ratios of
molecular to atomic being the attribute of early type galaxies (Young
\& Knezek 1989). We should point out that this relation has been
obtained without taking into account the effects of chemical
abundance and UV radiation field in the CO to H$_{2}$ conversion
factor. These effects have been demonstrated to be important in calculating
the molecular gas masses (see Verter \& Hodge 1995;
Wilson 1995).  Molecular gas radial distributions have exponential
shapes as a function of radius, mimicking the star formation rate
distributions.  Irregular galaxies, with lower mass, luminosity, and
absolute abundances, and exponential radial distributions for gas and
star formation rate follow the same behavior as above.  Their only
inconsistent feature with respect to those trends is their flat radial
abundance distributions.

\section{CHEMICAL EVOLUTION MODELS OF DISK GALAXIES}

The multiphase chemical evolution model, developed by Ferrini et al.
(1992; 1994), has been applied to several spiral galaxies (Moll\'a,
Ferrini \& D\'{\i}az, 1996: hereafter MFD96) with different
morphological types. The model reproduces the observed correlations by
varying the characteristics infall rate and the cloud and star
formation efficiencies with the galaxy morphological type and/or the
arm class as defined by Elmegreen \& Elmegreen (1987). The observed
radial distributions of atomic and molecular gas, of oxygen abundances
and of the star formation rate are used as observational
constraints. We can infer from the previous simulations that a galaxy
less massive than NGC~300 and NGC~598, which have been modeled by
MFD96, would have evolve more slowly, because of lower efficiencies in
forming molecular clouds and stars; this would produce an exponential
radial distribution of atomic gas and a steeper abundance radial
gradient than those obtained in MFD96 for NGC~300 and NGC~598.  The
strong gradient would be due to the high star formation rate in the
central regions. Massive stars from this episode would eject $\alpha$
elements, increasing the oxygen abundance mainly in this zone, while
the rest of the disk would maintain a low value. The difference
between the central level of abundance and the value in the outer
region would give the appearance of a steep radial gradient.

An important parameter influencing the shape of the radial
distribution of abundances is the initial radial mass distribution, or
indirectly the total mass of the galaxy.  The radial mass profile,
$M(R)$, may be indeed different for galaxies with the same Hubble type,
allowing different abundance distributions or gradients.  Our
aim, in this work, was to verify whether the multiphase chemical
evolution model would actually predict a steep radial gradient for the
oxygen abundance in the galaxy NGC~1313 or could produce the {\it
observed} flat gradient using a range of different assumptions, while
reproducing the observed radial distributions the gas and star
formation.

Not yet included at present in our model is the presence of a
bar. NGC~1313 has a short, but rather strong bar (Martin
1995). Barred galaxies have a flatter radial gradient (Edmunds \& Roy
1993; Martin 1992; Zaritsky, Kennicutt, \& Huchra 1994; Martin \& Roy
1994, 1995; Ryder 1995; Roy 1996).  It has been shown that bars induce
large-scale mixing due to radial flows (Friedli, Benz, \& Kennicutt
1994; Friedli \& Benz 1995).  The radial flows produce a net
accumulation of mass in the inner regions of the disk, and may enhance
a central star formation. This results first in a flatter abundance
gradient for the outer disk and in a steepening of the gradient for
the inner disk; this behavior is a local and transient phenomenon
which lasts a few hundred millions years. On the longer timescale,
radial mixing dominates and the global gradient is made flatter than
before the bar formation.  The dynamical and morphological evolution
of low mass galaxies has not been studied yet by theoreticians because
of the difficulties in building realistic codes of disk galaxies where
the gas component dominates the mass (D. Friedli, private
communication).

The multiphase model uses as input the total mass radial distribution,
obtained from the rotation curve; the possibility of the rotation
curve changing with time or the presence of dynamical processes, such
as radial flows of gas, are not taken into account.  The model has
been already applied to three massive barred galaxies (Moll\'{a},
Hardy \& Beauchamp 1998), and we know that it is not possible to fit
the abundance data in the bar region with the multiphase approach. To
describe these zones, a chemical evolution model including radial
flows of gas must be developed. The galaxy NGC~1313 has a bar, but it
will be assumed to be small enough not to have an important effect.
This paper assesses the validity of this simple assumption.

\section{PROPERTIES OF NGC~1313 FROM THE MULTIPHASE MODEL}

We have applied the multiphase model to the disk galaxy NGC~1313
which is an isolated very late-type barred galaxy. Its morphological
type lies between irregular magellanic spiral (Sm) and the latest type
of {\sl normal} spirals (Sd).  We modeled the galaxy chemical
evolution as in MFD96. We assumed that the protogalaxy is a spheroid
of primordial gas. The total mass included in the spheroid and its
radial distribution are inferred from the rotation curve. The rotation
curve is derived from the H\,{\sc i} 21 cm line data of Ryder et
al. (1995); the maximum rotation velocity given by these authors is
$V_{max}\sim$ 110 km/s.  The rotation curve shows an apparent hump at
7 kpc, but the authors warn that it may be an artifact due to the high
inclination of the galaxy. We use their fit to the data - with a
smoother behavior -- to derive the total mass of the corresponding
spherical protogalaxy.
\begin{table*}
\caption{ NGC~1313 Model Input Parameters}
\begin{tabular}{clc}
\tableline\tableline
\noalign{\smallskip}
Name & Meaning & Value \\
\tableline
$\tau_{\rm 0}$ & Collapse Time Scale & 12 Gyr \\ 
$\epsilon_{\rm K}$ & Efficiency for the star formation 
in the halo & 0.009 \\
$\epsilon_{\rm \mu }$ & Efficiency for the cloud formation
 in the disk & 0.05\\
$\epsilon_{\rm H}$ & Efficiency for the spontaneous star 
formation in the disk & 0.003 \\
$\epsilon_{\rm a}$ & Efficiency for the stimulated star
 formation in the disk&  2.5$\times 10^{-8}$\\
\tableline
\end{tabular}
\end{table*}

The galaxy is divided into 10 concentric cylindrical regions 1 kpc
wide; a {\sl halo} zone and a {\sl disk} zone correspond to each
galactocentric step. As the gas begins to form stars in the halo, it
collapses onto the equatorial plane, forming the disk.  The collapse
proceeds at a rate that is inversely proportional to the collapse time
scale.  Due to its dependence on the total local mass surface density,
this timescale, $\tau (R)$, changes as $\tau \propto \sigma (\rm
R)^{-1/2}$ (Jones \& Wyse, 1983). We assume this radial dependence to
have an exponential form: $\tau (R)=\tau _0 e^{\rm{-R}/l}$, where $l$
is the disk scale length taken as 2.5 kpc (Ryder et al. 1995).  The
characteristic collapse timescale, $\tau_0$, depends on the total mass
of the galaxy through the relation $\tau _0\propto \rm M_9^{-1/2}T$
(Gallagher, Hunter \& Tutukov, 1984), where $M_9$ is the total mass of
the galaxy in 10$^9\ \msun$, and T is its age, assumed to be 13 Gyr.
We calculate the collapse timescale $\tau _0$ with the expression
$\tau_{0}=\tau_{\odot}(\rm M_{9,gal}/M_{9,MWG})^{-1/2} $, where the
subscript MWG refers to the Milky Way, and $\tau_{\odot}$ taken as 4
Gyr, is the collapse timescale used for the solar neighborhood region
(Ferrini et al. 1994).  Total masses in units of $ \rm M_{9,gal}$ are
calculated as $\rm M_{gal}= 2.32 \times 10^{5} R_{gal} V_{max}^{2}$
(Lequeux, 1983), with R$_{\rm gal}= 15$ kpc; V$_{\rm max}$ has the
value 110 km/s already quoted. The collapse time finally selected is
$\tau_{coll} = 12$ Gyr; it is normalized to a region equivalent to the
solar region, and is located at a galactocentric distance R$_0$ which
is at 2 kpc based on a value of 1.73 kpc for the effective radius of
NGC~1313 (Ryder et al. 1995).

The initial mass function (IMF) is that of Ferrini, Palla, \& Penco
(1990), and the nucleosynthesis prescriptions are the same as used in
other multiphase models: Woosley \& Weaver (1986) for massive stars;
Renzini \& Voli (1981) for low and intermediate mass stars and Nomoto,
Thielemann \& Yokoi (1984) and Branch \& Nomoto (1986) for type I
supernova explosion yields.  $^{14}\rm N$ production is calculated
with Maeder (1983) prescriptions.  The star formation in the halo is
assumed to be a Schmidt law proportional to the gas density (smooth
distribution of atomic gas) with $n=1.5$ as an exponent. The
proportionality coefficient has a non linear dependence on the radius
through the halo volume, for every zone or region, and on the
efficiency of the process, $\epsilon_{K}$, which is assumed constant
for every region and galaxy.

Star formation in the disk is considered as a two step process: first,
molecular clouds form from the uniform atomic gas; then stars are
created by cloud-cloud collisions (spontaneous star formation). 
\begin{figure*}
\vspace*{8cm}
% psfile=#1 vsize=#2 angle=#3 hscale=#4 vscale=#5 hoffset=#6 voffset=#7
\includegraphics{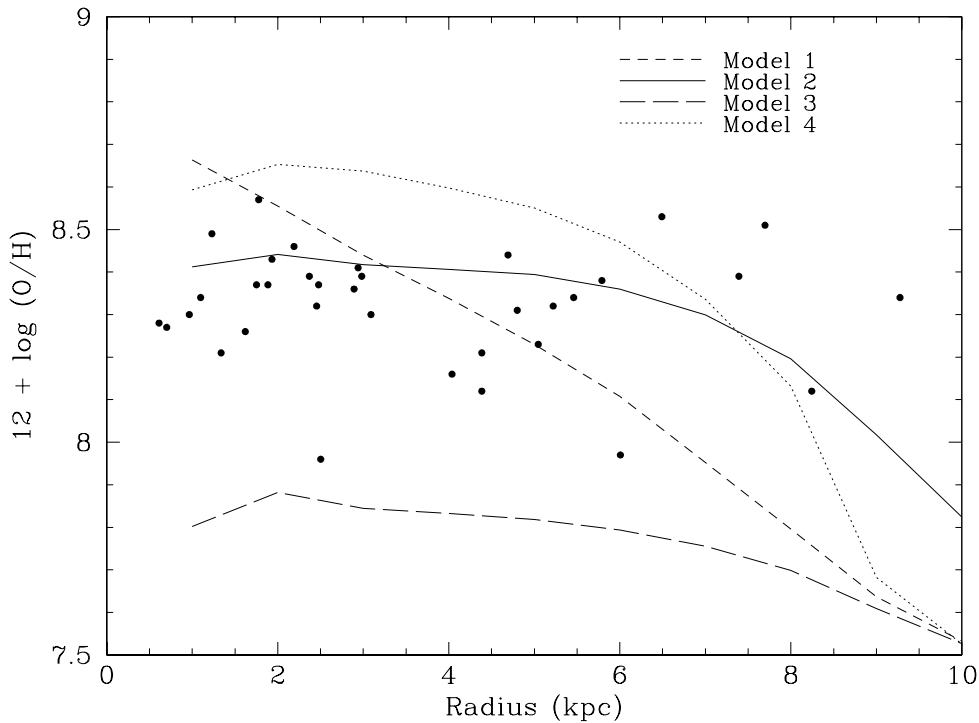}
\figcaption{
Predicted radial distribution of the oxygen abundance 
12 $+$ log (O/H) at the present time for NGC~1313:
Model (1) is represented by the short dashed line, model (2) by the
solid line, model (3) by the long dashed line, and model (4) by the dotted
line. Observational data points are from H\,{\sc ii} region
O/H abundance measurements by Walsh \& Roy (1997). }
\end{figure*} 
The induced star formation from the interaction of massive stars (through
their winds and/or supernova shells) with molecular clouds is also
included (stimulated star formation).  The observed star formation in
molecular clouds provides the empirical basis for our star formation
law, which may be closer to the true physical mechanism than other
assumed laws.  The star formation rate depends on the masses of the
different phases of the matter involved with proportionality
coefficients that depend on some efficiency or probability factors
(Table 1). 

 The induced star formation is proportional to the
efficiency, $\epsilon _a$, which is usually assumed constant for all
regions and galaxies.  The two other processes of formation have a
radial dependence due to the volume effect, and a dependence on the
efficiencies $\epsilon _\mu $ for the cloud formation and $\epsilon
_H$ for the stimulated star formation. The range for their
characteristic values is selected by taking into account the Hubble
type, and/or the arm class for the galaxy and the previous values for
all the Hubble types (see Table 4 of MFD96).  For NGC~1313, these
values are $\epsilon _\mu =0.05$ and $\epsilon _H= 0.003$ and are
applied to the equivalent  solar region with radius $\rm R_{0}$.

Final values for the efficiencies and collapse times have been
selected to construct a series of models fitting the observed radial
distributions of atomic gas (H\,{\sc i}), star formation rate surface
densities and oxygen abundance of the gas phase.  Observational data
for H\,{\sc i} surface densities are from Peters et al. (1994) and
Ryder et al. (1995). The star formation radial distribution is derived
from the H$\alpha$ flux radial profile of NGC~1313 by Ryder (1993);
we normalized it to the equivalent to solar region value $\Psi_{0}$.
Oxygen abundances are those of Walsh \& Roy (1997), who derived O/H
abundances using the calibration of the nebular diagnostic line ratio
([O\,{\sc ii}] + [{O\,{\sc iii}]/H$\beta$ by Edmunds \& Pagel
(1984). Unfortunately, there are no data on the molecular gas.

We ran several models trying to reproduce the observed radial
distributions in a self-consistent manner.  We show the results for
four different models in figures 1 to 4. These different models have
almost the same efficiencies $\epsilon_{\mu}$ and $\epsilon_{H}$ at
the distance equivalent to solar radius (2 kpc), but use different
assumptions for radial dependences, as we will explain in the next
paragraph.  Figure 1 show the predicted radial abundance distribution
compared with the corresponding observational data; the black dots are
the H\,{\sc ii} region O/H measurements of Walsh \& Roy (1997).
Figure 2 displays the theoretical radial distributions of the star
formation rate surface density, normalized to the value $\Psi_{0}$;
the actual rates, from the azimuthally averaged H$\alpha$ flux, are
shown by the black dots.  In Figures 3 and 4, the inferred radial
distributions of atomic H\,{\sc i} gas and molecular H$_{2}$ surface
densities are shown along the true data as squares or black dots for
H\,{\sc i} from Peters et al. (1994) and Ryder et al. (1995),
respectively.

\begin{figure*}
\vspace*{8cm}
% psfile=#1 vsize=#2 angle=#3 hscale=#4 vscale=#5 hoffset=#6 voffset=#7
\includegraphics{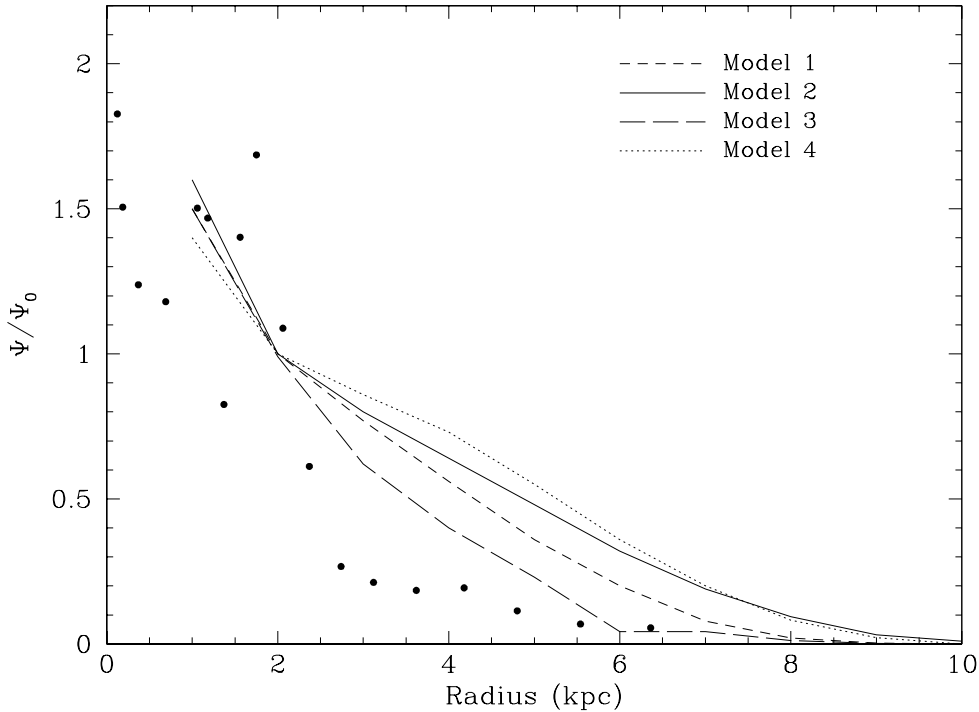}
\figcaption{
Predicted radial distribution of the star formation rate surface
density, normalized to the value $\Psi_{0}$. Observational data are from
H$\alpha$ flux measurements by
Ryder (1993). The lines referring to different models have the same meaning
 as in Figure 1.}

\end{figure*}

Model 1 is the {\sl standard} multiphase model.  The efficiency to
form molecular clouds $\epsilon_{\mu}$ varies with galactocentric
radius, being highest for the inner regions of the disk. The
efficiency $\epsilon_{H}$ to form stars from cloud-cloud collisions
process is kept constant with the radius. This model (1) is shown in
Figures 1 to 4 with the dashed line.  The atomic gas distribution
reaches a maximum at 4 kpc (Figure 3); the O/H gradient is very steep
($\sim -0.128 $ dex/kpc) (Figure 1). This is an expected result in
light of the sequence of models previously run for several other
galaxies. Actually, this model is very similar to the models of MFD96
for NGC~300 and NGC~598 using the same multiphase approach; NGC~300
and NGC~598 are observed to have maxima of H\,{\sc i} in the inner
disk (at $\sim$ 3 kpc ) and steep radial O/H gradients.  However, NGC~
1313 is not observed to have these features.  We need a model giving
an exponential atomic gas distribution, and at the same time, a flat
radial abundance gradient.  Though successful for two galaxies of
similar mass and luminosity, our calculations fail to adjust the
relevant physical distributions in NGC~1313.

Our assumptions for Model 1 do not appear valid.  We have supposed
that the efficiency to form molecular cloud, $\epsilon_{\mu}$, is
larger in the inner disk due to the effect of the spiral density
wave. Taking into account that NGC~1313 is more irregular in its
optical appearance than NGC~300 and NGC~598, the two galaxies of
comparison, we may relax this assumption and check whether a model
with an efficiency $\epsilon_{\mu}$ constant along the radius gives a
better result.  

We also assumed that once clouds are formed, the
efficiency to form stars from molecular is constant. But, with this
assumption, the observed exponential function of the star formation
rate was not fitted for NGC~300 nor NGC~598, because the distribution
had the tendency to flatten in the inner disk. To alleviate this
problem, recent models (Moll\'{a}, Hardy \& Beauchamp 1998) have been
computed with smaller efficiencies at shorter radii. The reason for
this variation is that gas densities need to be above a threshold
density for star formation to proceed (Kennicutt 1989).  This critical
density depends on the dispersion velocity, usually constant along the
disk in spiral galaxies, and on the epicyclic frequency. Although the
velocity dispersion usually does not vary much across the disk, it is
possible to find disk regions where it is larger. In normal galaxies,
these zones may be those where the bar, or the bulge, introduces a
strong perturbation in the velocity field.  Ryder et al. (1995) have
measured dispersion velocities in NGC~1313, where they average 14
km/s; they are larger than in normal galaxies.  They also found that
the velocity dispersion reaches a value of 25-30 km/s near the center
of NGC~1313.  If star formation efficiencies is higher where the
dispersion velocity is lower, stars should form relatively more easily
in the outer disk.  This would be consistent with our hypothesis that
it is more difficult to form stars in the center of NGC~1313 than in
the comparison galaxies.

Both these assumptions has been used to construct Model 2, shown by
solid lines in Figures 1 to 4. Due to the larger efficiencies to form
stars further away in the disk, we are successful at getting a flat
O/H radial distribution.  The radial distribution of the atomic gas is
also satisfactorily reproduced; the molecular gas forms in the same
proportion at all radii, by exhausting the atomic gas in the outer
regions of the disk, where the total disk mass is already small. Where
Model 2 fails is in the large star formation rate predicted in the
outer radii in comparison with the Model 1. A flat radial gradient of
abundances is obtained, but the model does not reproduce the rate
inferred from the observed H$\alpha$ profile (Figure 2): the predicted
star formation rate is flatter than observed, with unrealistically 
high values at larger radii. 

\begin{figure*}
\vspace*{8cm}
% psfile=#1 vsize=#2 angle=#3 hscale=#4 vscale=#5 hoffset=#6 voffset=#7
\includegraphics{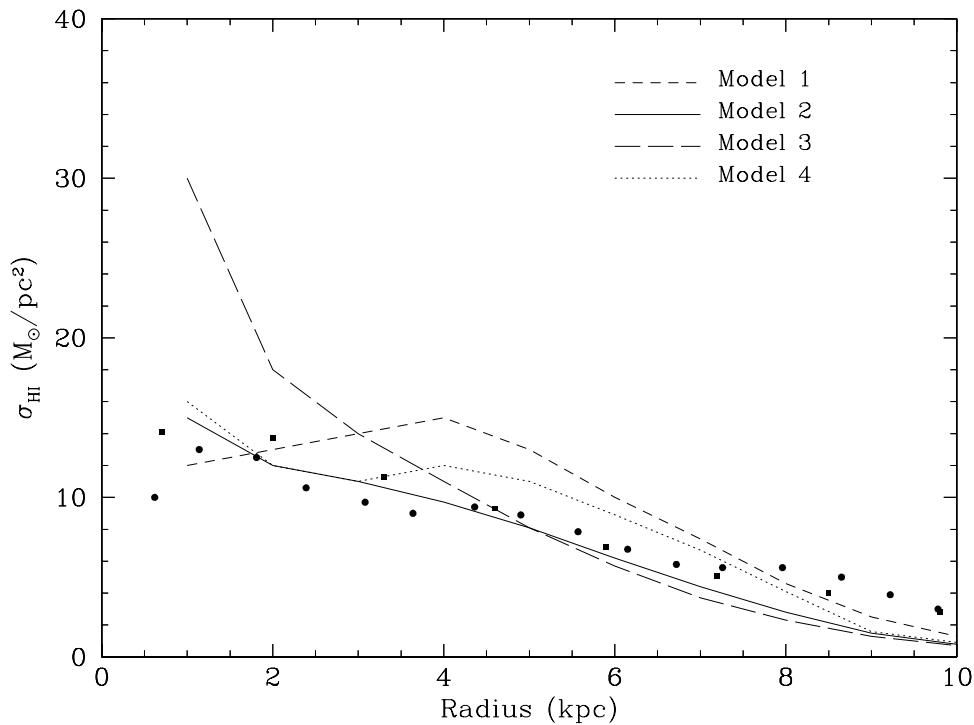}
\figcaption{
Predicted radial distribution of the atomic gas H\,{\sc i} surface 
density. The observed  data are taken 
from  Ryder et al.  (1995) -- circles -- and Peters et al. (1994) -- squares. 
The lines referring to different models have the same meaning
 as in Figure 1.}
\end{figure*}
Both these assumptions has been used to construct Model 2, shown by
solid lines in Figures 1 to 4. Due to the larger efficiencies to form
stars further away in the disk, we are successful at getting a flat
O/H radial distribution.  The radial distribution of the atomic gas is
also satisfactorily reproduced; the molecular gas forms in the same
proportion at all radii, by exhausting the atomic gas in the outer
regions of the disk, where the total disk mass is already small. Where
Model 2 fails is in the large star formation rate predicted in the
outer radii in comparison with the Model 1. A flat radial gradient of
abundances is obtained, but the model does not reproduce the rate
inferred from the observed H$\alpha$ profile (Figure 2): the predicted
star formation rate is flatter than observed, with unrealistically 
high values at larger radii.

In Models 3 and 4, variation in the relative importance of stimulated
star formation was tried. This is done by changing the parameter
$\epsilon_{a}$ which measures the rate of interaction between massive
stars and the surrounding molecular clouds. Normally, this parameter
is assumed constant, because this it is presumed to be a local process
(the star formation rate depends on the local gas surface density)
restricted to a relatively small region. However, this process
probably depends on the ultraviolet luminosity and on the density of
the gas and dust that may shield the region from the radiation field.
To explore the effect of enhanced stimulated star formation, we
decreased the efficiency of spontaneous star formation to let
stimulated star formation dominate.  If this quantity has a radial
distribution with an exponential function, the star formation rate can
be adjusted with this kind of model.  Model 3, having a low efficiency
$\epsilon_{a}$, leads to an exponential behavior for the star
formation rate and a flat abundance gradient, but the level of
abundance, $ 12 +\rm log (O/H) \sim 7.7$, is far below the observed
level. If we increase the star formation rate to raise the abundance
level by increasing the efficiency $\epsilon_{a}$ or the efficiency
$\epsilon_{h}$ -- this is Model 4 --, the star formation rate becomes
again unrealistically large in the outer parts.

\begin{figure*}
\vspace*{8cm}
% psfile=#1 vsize=#2 angle=#3 hscale=#4 vscale=#5 hoffset=#6 voffset=#7
\includegraphics{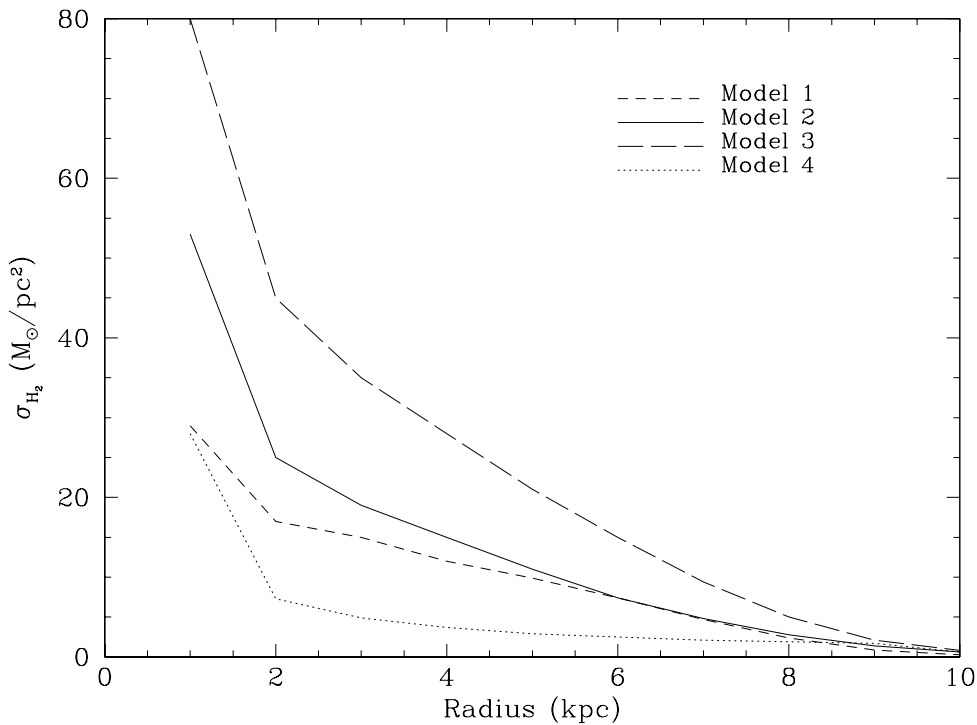}
\figcaption{
 Predicted radial distribution of the molecular gas H$_{2}$ surface 
density. The model lines have the same meaning as in Figure 1.}
\end{figure*}

\section{DISCUSSION AND CONCLUSIONS}

We tried to reproduce the radial abundance distribution
observed in the ionized interstellar gas of the galaxy NGC~1313,
which has a flat abundance gradient; we compared with models for two
similar disk galaxies, NGC~300 and NGC~598, showing a very steep
radial distribution of atomic gas.  We failed for NGC~1313, because
we were unable to fit the present observed star formation rate radial
distribution in a self-consistent way with other related
distributions, those of gas and of O/H. Taking into account the
observed global radial abundance gradients, the radial distributions
of star formation rates and of gas surface densities, we predicted
very steep radial O/H distributions for all three galaxies; the
observed distributions for atomic gas and star formation rate follow
our expectations, but that of O/H does not in NGC~1313.

Edmunds \& Roy (1993) remarked that global O/H radial gradients seem
to disappear when the spiral structure no longer exists.  Exploring
further this idea, we ran a model with a constant and low value for
the efficiency of molecular cloud formation and a small value
(decreasing in the inner regions) for the efficiency of spontaneous
star formation, to try to get a flat gradient in the radial abundance
distribution. It did not work: a flat gradient was obtained, but the
predicted star formation radial profile was much flatter than
observed.

On the other hand, the star formation rate being highest at the center,
we expect the oxygen abundance to be maximum there, as a
consequence of enhanced production of elements. Could then the
effective enrichment be less efficient because of loss of metals, thus
resulting in a flatter gradient? Various possibilities for element
loss exist.  Tenorio-Tagle (1996) has already suggested that recent
heavy element enrichment is unobservable because the fresh ejecta are
hiding in a hot phase.  A second possibility involves mass loss due to
collective winds produced by supernova explosions (Charlton \&
Salpeter 1989). The mass ejected expel $\alpha$-elements produced in
the first stellar generations out of the disk. This hot gas condenses
later and falls back over a long time-scale, resulting in an
homogeneous oxygen abundance distribution in the interstellar gas.
Actually, both possibilities may combine: supernova explosions produce
a galactic wind which ejects a proportion of elements already created.
The hot gas diffuses away from its injection location and eventually
falls on any region of the disk.  It eventually mixes with the cold
phase, and the elemental abundances seen in H\,\sc ii} regions show a
uniform radial distribution. This possibility is consistent with the
calculations of Hunter, Elmegreen \& Baker (1998) in their study of
star formation mechanisms in irregular galaxies. The authors suggest,
as a criterion for the onset of star formation, a critical density
allowing the existence of cool gas; this critical density
$\Sigma_{th}$ depends on the total pressure and on the average stellar
radiation field. They found that most galaxies have
$\Sigma_{g}/\Sigma_{th} \leq 1$, which suggests that the gas is in the
warm phase in irregular galaxies due to the lower pressure.  It means
that it is hard to sustain a cool phase in the H\,{\sc i} phase, and
that most of the gas is warm.

A big caveat is in order: why would the loss mechanisms work
differently in NGC~1313, as compared to NGC~598 or NGC~300 which have
about the same total mass?  Could this be due to a different merger
history?  The supershell found by Ryder et al. (1995) in NGC~1313
might be explained as the effect of a merger. However Ryder et al.
demonstrate that the shell must originate from an internal
event. Furthermore there is no indication from the H\,{\sc i} velocity
field nor from the photometry of the disk that a merger event took
place in the recent history of NGC~1313.  Thus the possibility of a
merger as the cause of the flat gradient is negligible.

Furthermore, flat gradients or uniform abundance distributions are a
normal feature of low-mass irregular galaxies, not an exceptional
characteristics. We must find a general mechanism producing a flat
oxygen abundance distribution, and giving at the same time consistent
corresponding radial distributions for star formation and atomic gas
surface densities.  This mechanism should also work for other low mass
irregular galaxies such as those in Table 3 of Walsh \& Roy (1997).
From a statistical point of view, the action of radial flows induced
by bars as the cause of flat gradients is more probable than rare
merger events.  Radial flows of gas appear as a consequence of the
non-axisymmetric gravitational potential; these radial flows are
efficient at mixing the interstellar medium and can produce a flat
gradient over a timescale of $\sim$1 Gyr (Friedli, Benz, \& Kennicutt
1994; Edmunds \& Greenhow 1995).

In the case of the galaxy NGC~1313, several authors (Blackman, 1981;
Marcelin \& Athanasoula 1982; Ryder et al. 1995) 
came to the conclusion that the  velocity field pattern could only
be explained by introducing a radial velocity component. This is
typically the case for barred galaxies;
the radial component may be as small as 20 km/s or as large as
80 km/s. For galaxies with bigger bars than NGC~1313,
radial flow velocities are  $\sim 40$ km/s (Blackman 1981; Ball
1992). For the Large Magellanic Cloud which is smaller
than NGC~1313, the recent work of Kim
et al. (1998) reveals  streaming motion
in the vicinity of the stellar bar; from these data,
we infer the amplitude of the perturbation
in the velocity field to be $\sim 10-15$ km/s.
Exact calculation of the velocity of radial flows
from the observed H\,{\sc i} velocity  field requires a complex numerical
simulation code; however by scaling for galaxy mass and
bar strength, we  estimate that the radial component due to a bar
may be $15-20$ km/s for NGC~1313.  Assuming  20 km/s, chemical mixing 
of the gas may be achieved
in a timescale between 0.5 and 2 Gyr following the
analysis of Martin \& Roy (1995) and Roy \& Kunth (1995). The shortest timescale  is obtained  by simply dividing the radius of the disk
by the mean radial flow velocity; however this is too simplistic.
The longest timescale is obtained by calculating the pseudo-diffusion
timescale $t= R^2/(v\ l)$, where $R$ is the radius
of the disk, $v$ the mean radial flow, and $l$ is the
width of the annuluses over which the net radial
flow is either inward or outward (see Fig. 2 in Friedli, Benz, \&
Kennicutt 1994). This upper value is probably the most
plausible, and is much shorter than the Hubble time.
Thus radial flow should influence the radial abundance distribution
if the galaxy is several Gyr old. Also with
radial flows, a large quantity of gas would accumulate in the center of the galaxy, and could generate a {\sl starburst}. This is consistent with 
soft X- ray
emission from the center of NGC~1313,  which could be produced by a
low-luminosity starburst or AGN (Colbert et al. 1995).

Because of mass conservation, radial flows 
must be compensated by the formation of stars in the
center which acts as a sink. If not, the disk will deviate from
a $1/r$ profile and will not be
stable (Struck-Marcell 1991). Massive star formation will insure
that the excess of mass is expelled out of the region through
some sort of wind phenomenon
similar to that produced by supernovae explosions. 
Maintaining a $1/r$ profile imposes a limit on the
radial flow velocity; this limits depends
on the  total local surfacic mass density and on the star formation rate in
the central region (Struck-Marcell 1991). We have calculated this limit to be
$v_{r} = 0.2 $ km/s for NGC~1313.  
If the bar induces a radial flow with a
velocity larger than $\sim$1 km/s, there should be some mass
loss. Therefore, both radial flows and mass loss by winds must be
incorporated into chemical evolution models.  We also need a multi-zone model
to calculate radial dependences and exchanges between radial zones. We hope to show in a near future
results from models with these characteristics. 

\begin{acknowledgements}

We had useful discussions with Daniel Friedli and Yvan Dutil.
The referee made very helpful suggestions.
M. Moll\'{a} acknowledges the Spanish {\sl Ministerio de Educaci\'{o}n y 
Cultura} for its support through a post-doctoral fellowship.
This investigation was funded in part by the Natural Sciences
and Engineering Research Council of Canada and by the Fonds
FCAR of the Government of Qu\'ebec.

\end{acknowledgements}

 \end{document}